\documentclass[letterpaper, 10 pt, conference]{ieeeconf}  
\IEEEoverridecommandlockouts                              

\usepackage{amsmath} 
\usepackage{amssymb}  
\usepackage{algorithm}
\usepackage{bm}
\usepackage{authblk}
\usepackage{cases}
\usepackage{graphics} 
\usepackage{graphicx}
\usepackage{epsfig} 
\usepackage{times} 
\usepackage{balance}
\usepackage{color}
\usepackage{epstopdf}
\usepackage[utf8]{inputenc}

\usepackage{booktabs}
\usepackage{mathrsfs}
\usepackage{theorem}

\usepackage{algorithmic}
\usepackage{tabularx}

\usepackage[hidelinks]{hyperref}
\usepackage{float}
\usepackage{makecell}
\usepackage{empheq}
\usepackage{xcolor}

\newtheorem{remark}{Remark}

\newcommand{\Shijie}[1]{\textcolor{black}{#1}}

\title{\LARGE \bf
Alternating Direction Based Sequential Boolean Quadratic Programming Method for Transmit Antenna Selection
}

	\author{Shijie Zhu$^{\dag}$, Xu Du$^{\dag}$

\thanks{\dag The first two authors contributed equally.
	SZ
	and XD
	are with the School of Information Science and Technology, ShanghaiTech University, Shanghai, China. 
	SZ is also with Innovation Academy for Microsatellites, Chinese Academy of Sciences, Shanghai, China.
	XD is also with Shanghai Institute of Microsystem and Information Technology, Chinese Academy of Sciences.
	SZ and XD are also with the University of Chinese Academy of Sciences, China while XD is with the Center for Intelligent Networking	and Communications, the National Key Laboratory of Science and Technology on Communications, University of Electronic Science and Technology of China.
	{\tt \{zhushj, duxu
		\}@shanghaitech.edu.cn }
}%
}
	
\begin{document}

\maketitle
\thispagestyle{empty}
\pagestyle{empty}

\begin{abstract}
The wireless mobile communication system is updated and iterated on the whole almost every decade. It is now in the development period of the application scenarios of the fifth generation mobile communication system (5G). Unfortunately, 5G relies on plenty of small base stations with a large number of antennas that consume a lot of energy. In this paper, a novel Boolean variable quadratic programming algorithm is designed 
for
the antenna selection optimization problem 
to reduce power consumption. Experiments show that the proposed algorithm achieves high complementarity satisfaction accuracy with only a few steps.


\end{abstract}

\section{INTRODUCTION}
With the explosion of wireless devices and services, challenges like climbing demand of communication capability cannot be solved easily. 
Fortunately, there are various promising technologies for 
5G
such as massive multiple-input multiple-output (mMIMO),
  turbo code etc \cite{wang2014cellular}.

 Among these, mMIMO is considered a leading technology
%
  since multiple antenna technology provides not only larger communication capacity with spatial multiplexing but reliability with spatial diversity \cite{tse2005fundamentals}, therefore
mMIMO with typically tens of antennas can address the aforementioned challenge
 \cite{wang2014cellular}. 
However, the power consumption 
corresponding to the rising numbers of antennas at the base station is becoming one of the research hotpots of economical concerns \cite{bjornson2015optimal}.
The consumption of electric power on base stations with massive antennas which are always active contributes to over 70\%
  of the electricity bill for the cellular operators
\cite{wang2014cellular}.
Therefore, the research direction of \emph{green communication} has been well studied 
for balancing the communication quality and power consumption and become one of the core 
of 5G 
 .
Green communication contains many techniques like new architecture such as intelligent reflecting surface (IRS) \cite{wu2019intelligent}
or resource allocation techniques, namely hybrid beamforming and antenna selection \cite{zhang2020prospective}, different variations of efficiency optimization objective has been well studied by \cite{bjornsonmassive}.
A \emph{Group Lasso} formulation has been used for 
turning off selected transmit source by \cite{shi2014group}, in details, the green communication framework with cloud radio access network can be formulated into a joint subset selection and power minimization beamforming problem.
Another article happens to coincide with an approach to select the active antennas with radio-frequency (RF) chains
 by using relaxation of Boolean variables, exhaustive search, interior point method  \cite{mahboob2012transmit}. However, the methods proposed in the above 
 literatures suffer a slow convergence rate and 
 low accuracy 
 for Boolean variables in the selection problem.


Historically, Boolean optimization has been explored in various fields. An early theory literature was proposed by \cite{connelly1963algorithms}, a Boolean linear programming exercise has been well studied \cite{boyd2004convex}. Moreover, \cite{luo2010sdp} solves a Boolean quadratic programming problem with $\{-1,1\}$ variables in a semidefinite programming fashion. Theoretically, pruning and sparsity optimization techniques can be also considered as Boolean variable optimization problems such as \cite{kouhalvandi2021automated} and \cite{zhou2019stochastic}. The scaling and approximation methods of Boolean variables are also often mentioned \cite{zhou2019stochastic}.
Nowadays, a new type of problem named \emph{Linear Complementary Quadratic Programming} (LCQP) \cite{lcqp} appears which aims at solving optimal control problems with complementary dynamic constraints, the idea is to linearize a corresponding penalization term 
without destroying the structure of \emph{QP}.
More related application can be found \cite{Nurkanovic2021} and \cite{lcqp}.
However, this algorithm can only solve the problem with objective function of QP and 
complementarity constraints,
which makes the application scenarios of this method limited.

In the present paper, we propose an \emph{Alternating Direction Based Sequential Boolean QP} (AD-SBQP) algorithm to solve general mix-integer nonlinear problem in an SQP fashion. In some literatures, AD is also named as alternating optimization (AO). The idea is to split Boolean and continuous variables in different steps without destroying the core structure of LCQP. The proposed algorithm shows better performance on 
 economic objective, complementarity satisfaction and computation time with limit iteration while the comparison with state-of-art methods has been shown in a given case study.

The rest of this paper is organized as follows: Section \ref{sec:Basics of Wireless Communication Systems} reviews the basics concepts of wireless communications\Shijie{.}
Sections \ref{sec:LCQP} proposed the AD-SBQP algorithm. 
And the numerical result is shown in Section \ref{sec: results}. 
\section{Basics of Wireless Communication Systems} \label{sec:Basics of Wireless Communication Systems}
In this section, we reviewed the basics of wireless communication that includes additive Gaussian white noise wireless channel model, maximum transmission ratio as well as antenna selection problem. Assuming perfect channel state information at transmit side (CSIT)
 is acquired, the system model can be idealized without propagation error during the process of signal transfer.
\subsection{Wireless Channel}

\begin{figure}[H]
 	\centering
 	\includegraphics[width=0.32\textwidth,height=0.125\textheight]{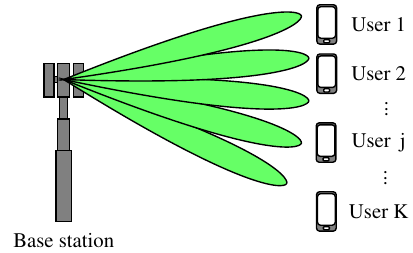}
	\caption{Multi-user MIMO communication network.}
 	\label{fig:mu-miso}
 \end{figure}

As shown in \autoref{fig:mu-miso}
 multi-user MIMO (MU-MIMO) is a system 
 equipped with multiple transmission and receiving antennas for broadcast information utilizing multipath propagation, while the operating bandwidth is defined as $B$ Hz. 
The wireless channel of a MIMO system
with $N$ transmit (TX) and $K$ receive (RX) antennas can be represented by a deterministic complex matrix 
\begin{equation}\label{eq:channel}
	\begin{split}
	H& 
	= [h_{(:,1)},h_{(:,2)},\cdots,h_{(:,j)},\cdots,h_{(:,K)}]
	\\& =
	 \begin{bmatrix} h_{11}& \cdots & h_{1K}\\\vdots & \vdots&\vdots \\ h_{N1}& \cdots & h_{NK}  \end{bmatrix}\in 
	 \mathbb C^{N\times K},  
	 \end{split}
\end{equation}
here channel gain $h_{ij}$ connects $i$th transmit antenna with $j$th user, $h_{(:,j)}$ indicates the vector including all $h_{ij}$ of $j$th user \cite{tse2005fundamentals}.

The relationship between the transmit side and receive side can be depicted as \cite{tse2005fundamentals}:
\begin{equation}\nonumber
	r = H^\top	t + n,
\end{equation}
here $r\in\mathbb{C}^{K}$, $t\in\mathbb{C}^{N}$ and $n$$\sim$${\mathcal{CN}(0,N_0\mathbf{I}_{K})}$ denote the received signal, transmitted signal and additive white Gaussian noise respectively. $N_0$ denotes the noise power density while $\mathbf{I}_{K}\in\mathbb{R}^{K}$ is an identity matrix.

The power allocation matrix 
\begin{equation}\label{eq:poweralloc}
	\begin{split}
		P &= 
		[p_{(:,1)},p_{(:,2)},\cdots,p_{(:,j)},\cdots,p_{(:,K)}]
		\\&=
		\begin{bmatrix} p_{11}& \cdots & p_{1K}\\\vdots & \vdots&\vdots \\ p_{N1}& \cdots & p_{NK}  \end{bmatrix}
		\in\mathbb{R}^{N\times K}
	\end{split}
\end{equation}
denotes the power amount associated with the channel gain. 
For instance, $p_{ij}$ represents the power allocated from 
$i$th
transmit antenna to 
$j$th
receive antenna, i.e. $p_{ij}$ correlates with the channel gain $h_{ij}$.


\begin{remark}
	To avoid ambiguity, we consider the multi-user multiple-input single-output (MU-MISO) as the special case of MU-MIMO when each user terminal only carries one antenna. 	
\end{remark}

As a kind of linear precoding schemes\footnote{\emph{Dirty paper coding} (DPC) is an optimal nonlinear precoding scheme which has high complexity especially with huge amounts of antennas while linear precoding can achieve 98\% performance of DPC \cite{gao2011linear}. 
}, the concept of maximum ratio transmission (MRT) was introduced in \cite{lo1999maximum} to maximize the signal-to-noise ratio (SNR) at each receiver (i.e. $j$th user) in multi-antenna communication,
  in details,
\begin{equation}\label{eq:mrtsnr}
\mathrm{SNR}_j = 
\frac{\sum_{i=1}^{N}p_{ij}\vert h_{(:,j)}^H v_j\vert^2}{N_0B}
\end{equation}
 with given beamforming directions \cite{bjornson2013optimal}
\begin{equation*}\label{eq:mrtprecode}
v_j = \frac{h_{(:,j)}}{\Vert h_{(:,j)}\Vert} ,\forall j .
\end{equation*} 
Here 
$(\cdot)^H$
denotes Hermitian transpose.

\subsection{Shannon Capacity}
In this paper \emph{Shannon Capacity} relies on both MRT and transmit antenna selection (TAS).
With \eqref{eq:mrtsnr},
\emph{Shannon Capacity} \cite{tse2005fundamentals} can be formulated as
\begin{equation}
	\begin{aligned}\label{eq:mrtcapacity}
		R(P)
		&=\sum_{j=1}^{K}B\log_2(1+\mathrm{SNR}_j)\\
		&=\sum_{j=1}^{K}B\log_2\left(1+\frac{\sum_{i=1}^{N}p_{ij}\frac{|h_{(:,j)}^Hh_{(:,j)} |^2}{\Vert h_{(:,j)}\Vert^2}}{N_0B}\right)\\
		&=\sum_{j=1}^{K}B\log_2\left(1+\frac{\sum_{i=1}^{N}p_{ij}\Vert h_{(:,j)}\Vert^2}{N_0B}\right)
	\end{aligned}
\end{equation}
which depicts the 
ideal
communication rate that the wireless communication system could get.
TAS is a signal processing method to 
save the cost of RF chains connected with transmit antennas equipped at base station (BS), when the set of antennas is selected with the principle of maximizing the downlink capacity \cite{sanayei2004antenna}. 

\begin{figure}[H]
	\centering
	\includegraphics[width=0.35\textwidth,height=0.25\textheight]{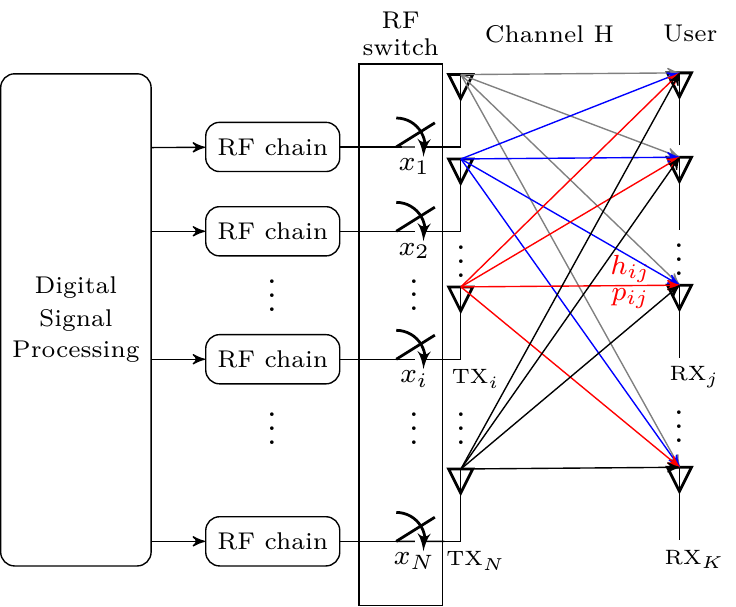}
	\caption{Multi-user MISO wireless communication system with transmit antenna selection.}
	\label{fig:abslinks}
\end{figure}

As shown in \autoref{fig:abslinks}, the switch vector $x\in\{0,1\}^{N}$ controls the on-off state of each RF chain with the corresponding transmit antenna \cite{gao2015massive}.
 By considering the switch variable $x$, \eqref{eq:mrtcapacity} can be reformulated into 
\begin{equation}
	\begin{aligned}\label{eq:capacity}
		\tilde{R}(P,x)
		&=\sum_{j=1}^{K}B\log_2\left(1+\frac{\sum_{i=1}^{N}p_{ij}x_i\Vert h_{(:,j)}\odot x\Vert^2}{N_0B}\right),
	\end{aligned}
\end{equation}
here Hadamard product $\odot$ indicates the element wise multiplication.
\subsection{Economic Sum Rate} \label{sec:ESR}
Inspired by \cite{shi2014group} and \cite{li2016dynamic} whose bi-linear structure with Boolean variable is nonconvex,
here we propose a simplified \emph{Economic Sum Rate} problem with the following linear summation objective function 
\begin{subequations}\label{eq:ecosumrate}
	\begin{align}
	\min_{ P, x}\;\;&  f(P,x)=\sum_{i=1}^{N}(\sum_{j=1}^{K}p_{ij}x_i+ p^{RF}x_i) \label{eq:specificobject}\\
	\quad\mathrm{s.t.}\;\;&
	\tilde{R}(P,x)
\geq R^{th}\label{eq:specific sum rate constraint}\\
		&x_i\in \{0,1\}, \qquad i = 1,2,\dots,N \label{boolean constraint}\\
	&\sum_{j=1}^{K}p_{ij}\le p^{th} \label{eq:per-antenna power constraint}\\
	&
	0
	\le p_{ij} ,\qquad\quad\dbinom{i = 1,2,\dots,N}{j = 1,2,\dots,K}\label{eq:per-user power constraint}
	\end{align}
\end{subequations}

Here, \eqref{eq:specificobject} can be viewed as a cost function while the sum rate inequality constraint \eqref{eq:specific sum rate constraint} is certainly satisfied with a given  threshold sum rate $R^{th}$.
 A constant $p^{RF}$ stands for the standby power of RF chain,
and a constant $p^{th}$ indicates the upper bound of power allocated at each transmit antenna.

\section{Alternating Direction Based Sequential Boolean Quadratic Programming}\label{sec:LCQP}
In this section, after basic theory of BQP is reviewed, an extension of which named \emph{Alternating Direction Based Sequential BQP} is proposed for solving general nonlinear programming problem with linear complementarity constraints by splitting normal and Boolean variables into different steps.

\subsection{Basics of BQP }
Consider a tuple $(\mathcal D, \mathcal Y, \mathcal Z)\in \mathbb{R}^{n_d}\times\mathbb{R}^{n_C\times n_d}\times\mathbb{R}^{n_C\times n_d}$, linear complementarity \cite{lcqp} can be defined as
 \begin{subequations}
 	\label{eq:totalcompcons}
 	\begin{empheq}[left ={0\leq Yd \perp Zd \geq 0\Leftrightarrow \empheqlbrace}]{align}
 		Yd &\ge  0\label{leftcompcons} \\
 		Zd  &\ge  0\label{rightcompcons} \\
 		d^\top Y^\top Zd&= 0,\label{bilinearcompcons}
 	\end{empheq}
 \end{subequations}
 $(d,Y,Z)\in (\mathcal D, \mathcal Y, \mathcal Z)$ collects the decision variable and linear transformation
matrix
 respectively. 
 Boolean constraint \eqref{boolean constraint} in \emph{Economic Sum Rate} \eqref{sec:ESR} can be somehow treated as a special case of \eqref{eq:totalcompcons} and can be expressed as
  \begin{subequations}
 	\label{eq:01compcons}
 	\begin{empheq}[left ={0\leq x \perp (\bm{1}-x) \geq 0\Leftrightarrow \empheqlbrace}]{align}
 		x &\ge  0\label{eq:leftcompcons} \\
 		\bm{1}-x  &\ge  0\label{eq:rightcompcons} \\
 		x^\top(\bm{1}-x)&= 0\label{01rcompcons}
 	\end{empheq}
 \end{subequations}
 where $\bm{1}$ represents a vector with all entries being one.

Based on above definition, \emph{Boolean Quadratic Programming} (BQP) can be depicted as the following equation
 by introducing Boolean 
 constraint \eqref{eq:01compcons}
\begin{subequations}\label{eq:boolCQP}
	\begin{align}
	BQP:\min_{x}\;\;&\frac{1}{2}x^\top Qx+g^\top x\label{lcqp:obj}\\
	\mathrm{s.t.}\;\;&Ax-u\leq 0 \label{lcqp:lpcons}\\	
	&0\leq x\perp(\bm{1}-x)\geq 0,\label{eq:lcqpbool}
	\end{align}
\end{subequations}
here $0\prec Q\in\mathbb{R}^{n_x\times n_x}$, $g\in\mathbb{R}^{n_x}$, $A\in\mathbb{R}^{n_A\times n_x}$ and $u\in\mathbb{R}^{n_A}$ collect the corresponding parameters. 
Here we introduce a penalty function \cite{ralph2004some} relate to \eqref{01rcompcons},  
\begin{equation}
\varphi(x) = x^\top (\bm{1}-x)\label{eq:penfcn}.
\end{equation}
Therefore, \eqref{eq:boolCQP} can be 
reformulated as penalty BQP (pBQP)
\begin{equation}
	\begin{split}
		pBQP:\min_{x}\;\;&\frac{1}{2}x^\top Q x +g^\top x+\rho\cdot \varphi(x)\label{rholcqp:obj}\\
		\mathrm{s.t.}\;\;&\eqref{eq:leftcompcons},\eqref{eq:rightcompcons},\eqref{lcqp:lpcons}
	\end{split}
\end{equation}
with a penalty parameter $\rho\in\mathbb{R}$.

The following BQP algorithm inherits the key idea of \cite{lcqp} by using linear approximation of the penalty function in Step \ref{step:app of LCQP}. 

\begin{algorithm}[H]
	\small
	\caption{BQP method}\label{alg:LCQP}
	\textbf{Input:}  coefficients $Q, g, A, u$, a termination tolerance $\epsilon>0$, an initial factor $\rho>0$ and update rate $\beta>1$.\\
\textbf{Repeat:} 
\vspace{0.35cm}
	\begin{enumerate}
		\item \emph{Globally Search:} solve QP without complementarity constraints:\label{step: global search}
		\begin{equation}
		\begin{split} 
		\hat x=\arg\min_{x}\;\;&\frac{1}{2}x^\top Q x+ g^\top x\label{step:globsearch} \\ \quad\mathrm{s.t.}\;\;
		&\eqref{eq:leftcompcons},\eqref{eq:rightcompcons},\eqref{lcqp:lpcons}
		\end{split}
		\end{equation}
		\item \emph{Penalty Function Approximate:}
		\begin{equation*}
		\begin{split}
		\varphi(x)&\approx \varphi(\hat x) + (x-\hat x)^\top \nabla \varphi(\hat x)\\
		&= (\varphi(\hat x) -\hat x^\top \nabla\varphi(\hat x))+x^\top \nabla\varphi(\hat x)
		\end{split}
		\end{equation*}
		
		\item \emph{Locally Search:} Minimize the reformulated penalty QP\footnotemark: 
		\label{step:app of LCQP}
		\begin{equation*}
			\begin{aligned} 
				\tilde{x}=\arg\min_{x}\;\;&\frac{1}{2}x^\top Q x +(g+\rho \nabla\varphi(\hat x))^\top x \\ \quad\mathrm{s.t.}\;\;
				&\eqref{eq:leftcompcons},\eqref{eq:rightcompcons},\eqref{lcqp:lpcons}
			\end{aligned}
		\end{equation*}
		\item  \emph{Line Search and Termination Criterion\footnotemark:} 
		
		$\alpha = StepLength(\hat x, \tilde{x}, \rho);$

		$\hat{x} \approx \hat{x} +\alpha (\tilde{x}-\hat x)$
		
		check if $\|\varphi(\hat x)\|\leq \epsilon$, if not, go to step \ref{step:update}.
		\vspace{0.35cm}
		\item \emph{Penalty Parameter Update:}\label{step:update}
		
		$\rho=\beta
		\cdot
		\rho$ 
		and return step \ref{step:app of LCQP}
	\end{enumerate}
\vspace{0.35cm}
\textbf{Output:} $x^*\leftarrow \hat x$.
\end{algorithm}
\footnotetext[2]{Complementarity $\varphi(x)$ \cite{lcqp} cannot be utilized due to the nonlinearity and nonconvexity properties, which can be replaced by the first order Taylor approximation $(\bm{1}-2\hat{x})^\top x$ that keeps the structure of QP.}
%
%
%
\footnotetext[3]{$\alpha$ can be obtained by any kind of line search method \cite{nocedal2006numerical}.
 }

\subsection{Theory of AD-SBQP}
For a generic nonlinear programming problem with Boolean constraints 
\begin{subequations}\label{eq:orinlp}
	\begin{align}
	\min_{x}\;\;&F(x)\label{eq:orinlpobj}\\
	\mathrm{s.t.}\;\;& c(x)\le 0 \;|\lambda\label{eq:oriratecons}\\
	&0\leq x \perp (\bm{1}-x)\geq 0\label{eq:orinlpbool}
	\end{align}
\end{subequations}
with smooth nonlinear maps $F:\mathbb{R}^n\rightarrow \mathbb{R},\ c:\mathbb{R}^n\rightarrow \mathbb{R}^m$. The Lagrangian function
 of which 
 is constructed as $\mathcal{L}(x,\lambda)=F(x)+\lambda^\top c(x)$ with a multiplier $\lambda\in\mathbb{R}^{m}$ of the corresponding dimension \cite{nocedal2006numerical}.
Assume $\bar{x}$ and $\bar{\lambda}$ are obtained from the previous nonlinear programming (NLP) iteration, a second-order Taylor expansion of the Lagrangian function $\tilde{F}(x)$ is shown as below with $\nabla^2\mathcal{L}(\bar{x}, \bar{\lambda})\succ 0$,
\begin{equation}
	\begin{aligned}
\tilde{F}(x)=
 &\frac{1}{2}(x-\bar{x})^\top\nabla^2\mathcal{L}(\bar{x}, \bar{\lambda})(x-\bar{x})\\
&+\nabla F(\bar{x})^\top
(x-\bar{x})+F(\bar{x})\label{slcqp:approx}.
\end{aligned}
\end{equation}
Inspired from the framework of Sequential QP (SQP), \eqref{eq:sLCQP} can be treated as a single step of \emph{Sequential Boolean QP} (SBQP) which connects with \eqref{eq:boolCQP} and \autoref{alg:LCQP}.
\begin{subequations}\label{eq:sLCQP}
	\begin{align}
	\min_{x}\;\;&\tilde{F}(x)\label{eq:sLCQPobj}
	\\
	\mathrm{s.t.}\;\;& \nabla c(\bar{x})(x-\bar{x})+c(\bar{x})\le 0\label{eq:affinecons}\\
	&0\leq x \perp (\bm{1}-x)\geq 0.\label{eq:01vectorcons}
	\end{align}
\end{subequations}
 
\begin{figure}[H]
	\centering
	\includegraphics[width=0.45\textwidth,height=0.35\textheight]{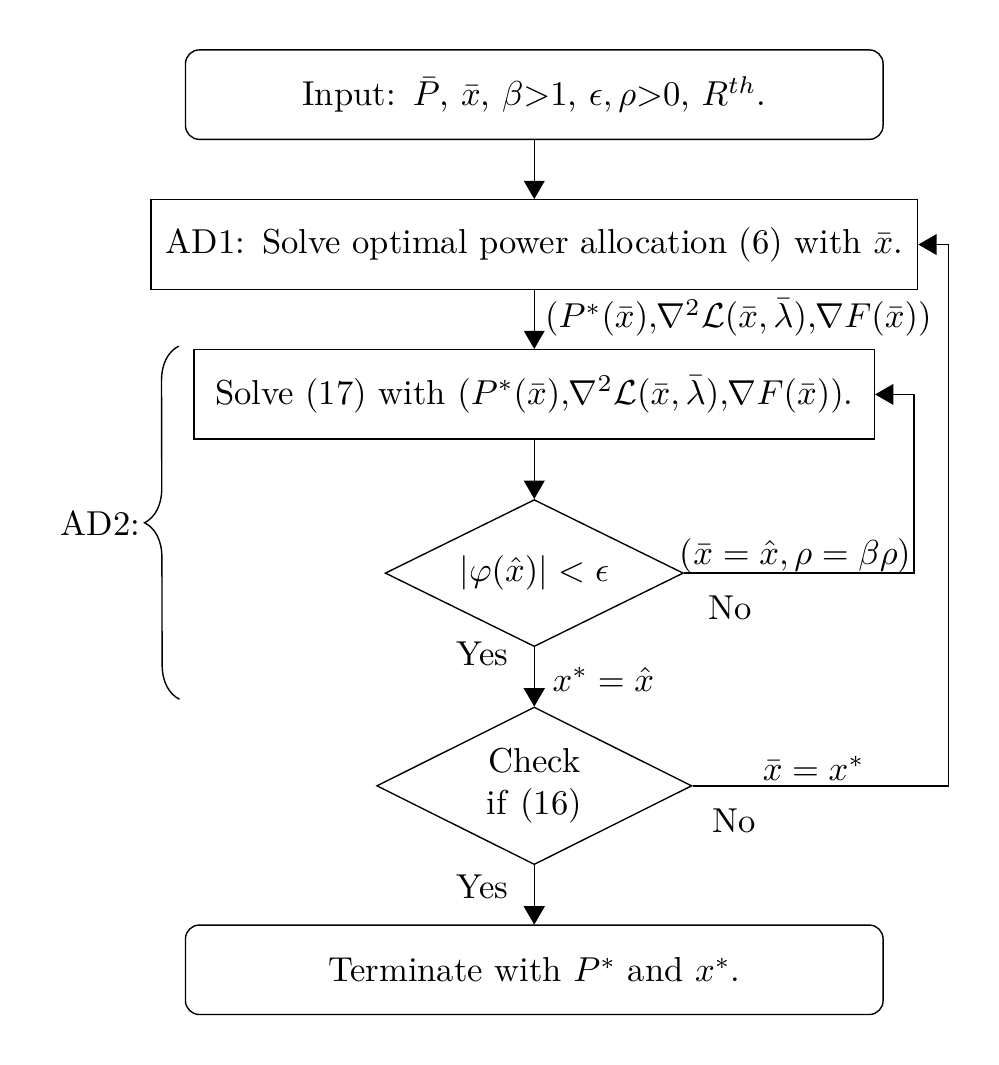}
	\caption{Flowchart of AD-SBQP.}
	\label{fig:adsbqpflow}
\end{figure}

The above discussion summarizes the solution of generic nonlinear Boolean optimization problem. When there are additional continuous variables in the optimization problem, the local optimal solution can be obtained by optimizing different kinds of variables in alternate directions without destroying the SBQP structure. With this spirit of \autoref{alg:LCQP}, \autoref{alg:AD-SLCQP} and \autoref{fig:adsbqpflow} show an new version of BQP named 
\emph{Alternating Direction Based SBQP method} (AD-SBQP).


\begin{algorithm}[H]
	\small
	\caption{Alternating Direction Based SBQP method}\label{alg:AD-SLCQP}
	\textbf{Input:}  
	initial guess of $\bar x\in \mathbb R^N$, $\bar P\in \mathbb R^{N\times K}$ a termination tolerance $\epsilon>0$, an initial factor $\rho>0$ and update rate $\beta>1$.\\
	\textbf{Repeat:}
	\begin{enumerate}
		\item \emph{Optimal Power Allocation (AD1):}\label{power allocation OPT} Solve \eqref{eq:ecosumrate}
	 with constant $\bar x$	by any NLP solver	
		Output: $P^*(\bar{{x}})$,
		 $\bar{\lambda}$
		 .
		\item \textit{Sequential BQP (AD2):}\label{alg:nonsmooth OPT} set
		\begin{equation*}
			\left\{
			\begin{aligned}
				F(x)&=f(P^*(\bar x),x)\\
				c(x)&=\tilde{R}(P^*(\bar x),x)
			\end{aligned}
		\right.
		\end{equation*}
		 solve \eqref{eq:orinlp} by SBQP \eqref{eq:sLCQP} and \autoref{alg:LCQP}.
		 
		Output: $x^*$
		\item \textit{Termination Criterion:}\\Check if 
		\begin{equation}\label{eq:termcri}
			\|
			\begin{bmatrix}
				P^*(\bar x)|x^* 
			\end{bmatrix}
			-
			\begin{bmatrix}
				\bar{P} |	\bar{x} 
			\end{bmatrix}
			\|
			\leq \epsilon
		\end{equation}

		if not go back to step \ref{power allocation OPT}) and set $\bar P= P^*(\bar x)$, $\bar x=x^*$.
	\end{enumerate}
\textbf{Output:} $P^*, x^*$.
\end{algorithm}
\subsubsection{AD1}
This stage, which is derived from \emph{water-filling} \cite{tse2005fundamentals}, aims to reduce the power consumption of the aforementioned MIMO system with fixed state $x$. The Quality of Service (QoS) requirement \eqref{eq:specific sum rate constraint}, the lower bound of $p_{ij}$ \eqref{eq:per-user power constraint}, and the upper bound of power for each antenna \eqref{eq:per-antenna power constraint} are all considered as physical constraints.
Moreover, the constraint \eqref{eq:per-antenna power constraint} as the upper bound denotes the power of each antenna, that ensures the optimized sum power will not exceed the preset upper bound.
\subsubsection{AD2}
After AD1, the Hessian $\nabla^2\mathcal{L}(\bar{x}, \bar{\lambda})$ and gradient $\nabla F(\bar{x})$ are evaluated jointly with $(P^*(\bar x),\bar \lambda)$.
Assume the local minimizer of \eqref{eq:sLCQP} is in the neighborhood of the global minimizer \eqref{step:globsearch},
In order to search a specific local minimizer, the objective function \eqref{eq:sLCQPobj} is replaced by
\begin{equation}
	\begin{aligned}
		\hat{F}(x)
		=\; &\tilde{F}(x)+\rho \nabla\varphi(\bar{x})^\top(x-\bar{x})\\
		=\; &\frac{1}{2}(x-\bar{x})^\top\nabla^2\mathcal{L}(\bar{x}, \bar{\lambda})(x-\bar{x})\\
		&+(\nabla F(\bar{x})+\rho \nabla\varphi(\bar{x}))^\top
		(x-\bar{x})+F(\bar{x})
		\label{slcqp:withpen}
	\end{aligned}
\end{equation}
which inherits the spirit of Step \ref{step:app of LCQP} in \autoref{alg:LCQP}. Moreover, another key part of AD2 is the complementarity tolerance check.
%

\section{Numerical Experiments} \label{sec: results}
In this section, we present the numerical performance of \autoref{alg:AD-SLCQP} drawing upon a MU-MISO TAS communication network shown in \autoref{fig:abslinks}. We illustrate the numerical comparison among 4 different configurations of wireless network.

\subsection{Data and Environment}
\begin{figure}[H]
	\centering
	\includegraphics[width=0.4\textwidth,height=0.125\textheight]{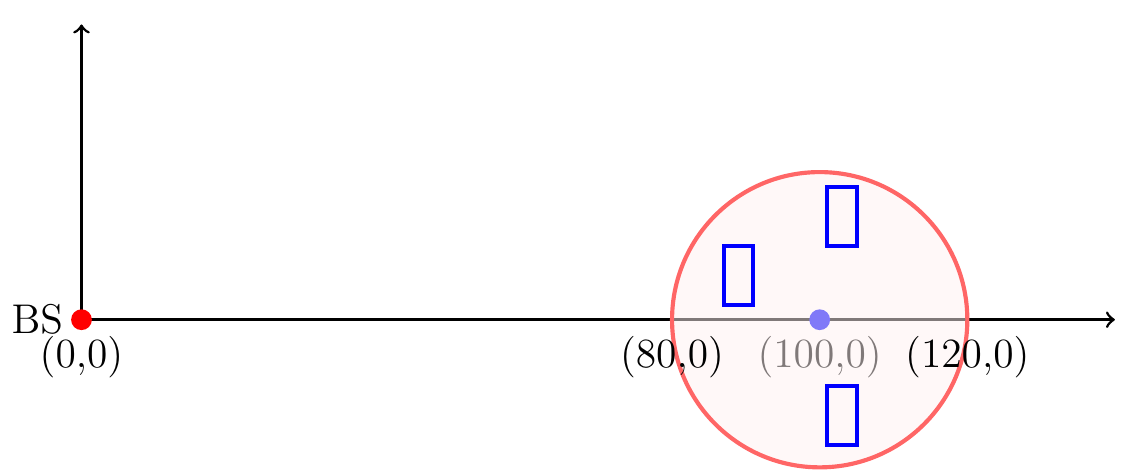}
	\caption{A two-dimensional Cartesian coordinate system: red solid point in origin denotes the BS while blue rectangles in red ring denotes users.}
	\label{fig:commmodel}
\end{figure}

In a two-dimensional Cartesian coordinate system, a  wireless communication network is shown as \autoref{fig:commmodel}: a) a BS at the origin point equipped with a uniform linear array of $N=64$ antennas, b) $K=64$ users are distributed randomly in a circular cell centered at $(100,0)$ with a radius of 20 meters. $\xi(\delta_j)=T_0\delta_j^{-\eta}$ denotes the distance-dependent path loss with parameters c) the distances $\delta_j$ between $j$th user and the BS , d) the path loss at reference distance per meter $T_0=-30$ dB and e) the path loss exponent $\eta=3.67$ \cite{access2010further}. We denote $h_{(:,j)} = \sqrt{\xi(\delta_j)}\tilde{h}_{(:,j)}$ where $\tilde{h}_{(:,j)}$ denotes $j$th origin channel gain subjects to Rayleigh fading due to the rich environment of scattering.
In order to simplify the experiment, the total power is normalized to 1 while the noise $N_0B$ at each user is also normalized to 1.
The $p^{RF}$ is set as $0.0078$ while the threshold sum rate $R^{th}$ is set as $82.71$.
The implementation of SBQP relies on \texttt{qpOASES} \cite{qpoases} while other parts depend on \texttt{IPOPT} \cite{ipopt}.
Both subsolvers are utilized through the \texttt{Casadi} \cite{casadi} interface with \texttt{MATLAB 2021b}.

\subsection{Implementation and Numerical Comparison}
Note that the AD framework consists of two essential parts: AD1 can be solved by any NLP solver,
while AD2 is a nonconvex and nonsmooth problem.
In AD2, we set $\rho=1$ as initial penalty parameter and its update factor $\beta=2$. 
Moreover, we select Armijo rule as line search method. 
We initialize the switch vector $\bar{x}$ as $\frac{1}{2}\cdot$$\bm{1}$ to ensure the fairness between the lower and upper bound. To assess the numerical effect of the proposed AD-SBQP algorithm, we compare \autoref{alg:AD-SLCQP} with other two variants of penalty BQP: one method, named by NSPen, solves \eqref{eq:orinlp} by adding \eqref{eq:penfcn} into \eqref{eq:orinlpobj} to substitute for \eqref{eq:orinlpbool}, the other penalty method, SPen, solves \eqref{eq:sLCQP} with the same penalty function \eqref{eq:penfcn}. Therefore, we call the other two methods in same AD framework as AD-NSPen and AD-SPen temporarily.
\begin{table}[h]
	\caption{Numerical Comparison}
	\label{table:comparison}
	\begin{center}
		\begin{tabular}{|c||c|c|c|}
			\hline
			Methods& AD-SBQP & AD-SPen & AD-NSPen\\
			\hline
			objective & 0.5269 & 0.5933 & 0.6093\\
			\hline
			complementarity & 2.9816e-19 & 6.4000e-07 & 6.4000e-07\\
			\hline
			time (seconds) & 281.4971 & 519.0302 & 1.3190e+03\\
			\hline
		\end{tabular}
	\end{center}
\end{table}

\autoref{table:comparison} shows the comparison among objective function, complementarity satisfaction and operation time. It was apparent that the objective function obtained by AD-SBQP is the lowest. Moreover, one can see that the precision of AD-SBQP in complementartiy is higher than other algorithms while its computation time is much less. 

\begin{figure}[H]
	\centering
	\includegraphics[width=0.355\textwidth,height=0.21\textheight]{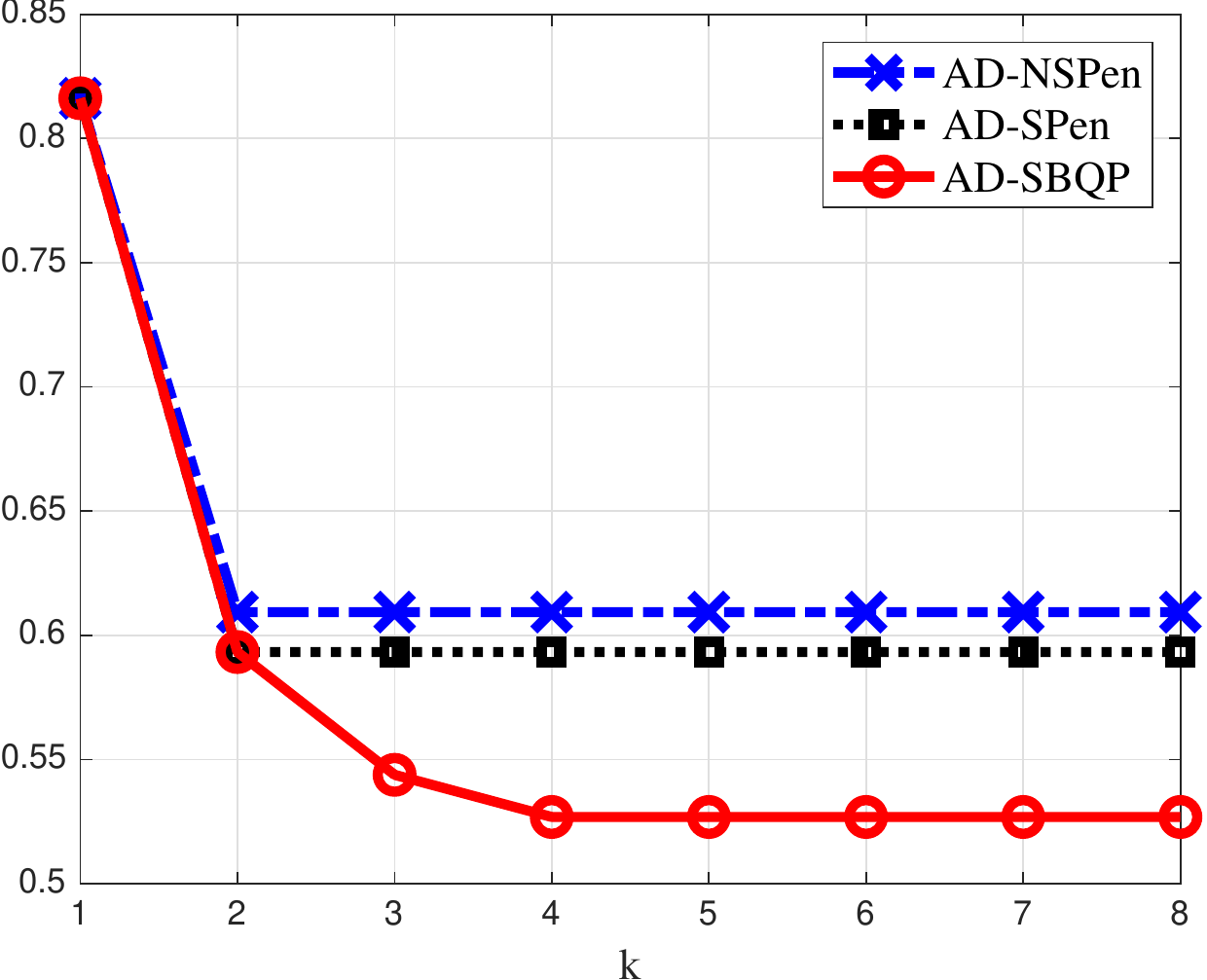}
	\caption{The blue dash-dotted and black dotted line represent the objective functions with AD-NSPen and AD-SPen respectively. The red solid line denotes the numerical result optimized by \autoref{alg:AD-SLCQP}.}
	\label{fig:objfcneach}
\end{figure}

\autoref{fig:objfcneach} shows the objective convergence comparison by using the above three methods, all of which converge in four steps. Note that  AD-NSPen and AD-SPen solve optimization problem with \eqref{eq:lcqpbool} directly and trap into different local minimizer while Step \ref{step: global search} of \autoref{alg:LCQP} avoid it by optimizing without $\varphi(x)$ term.




\begin{remark}
	When the complementarity tolerance is set as $10^{-10}$, even the penalty parameter $\rho$ increase to a very large value as $2^{32}$ of AD2, both AD-SPen and AD-NSPen will not well satisfy the complementarity tolerance.
\end{remark} 

\begin{figure}[H]
	\centering
	\includegraphics[width=0.475\textwidth,height=0.31\textheight]{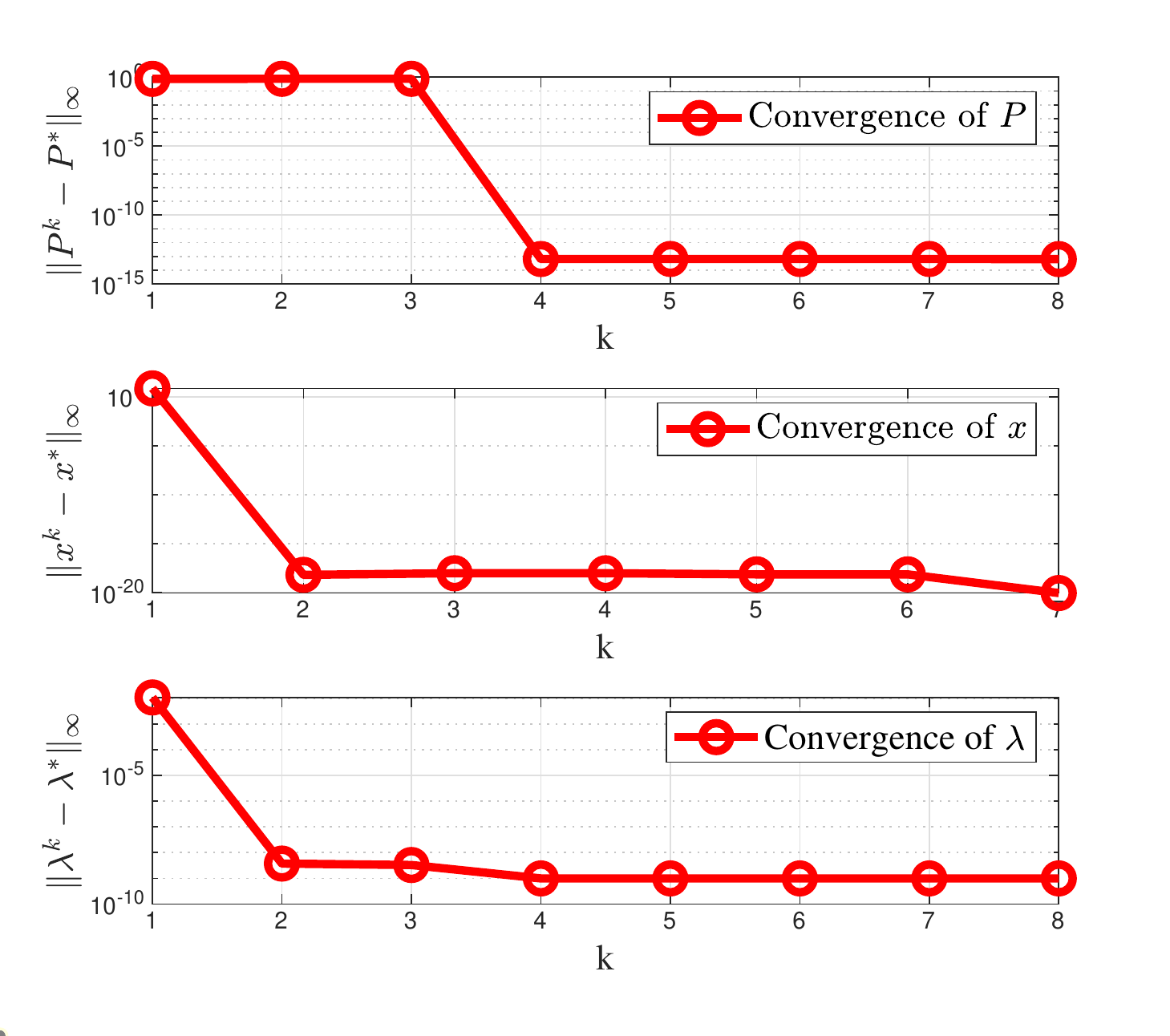}
	\caption{Convergence of primal variables $(P,x)$ and dual $\lambda$.}
	\label{fig:converglcqp}
\end{figure}

\autoref{fig:converglcqp} shows the convergence of primal $(P,x)$ and dual variables $\lambda$. One can see that all of the variables converge to optimal solution with only four steps, thus the proposed method can be potentially treated as a real time algorithm. 

Furthermore, the objective convergence of $8\times 8,16\times 16,32\times 32$ antenna configuration are similar to \autoref{fig:objfcneach} and \autoref{fig:converglcqp} with less operation time.

\begin{figure}[H]
	\centering
	\includegraphics[width=0.4\textwidth,height=0.24\textheight]{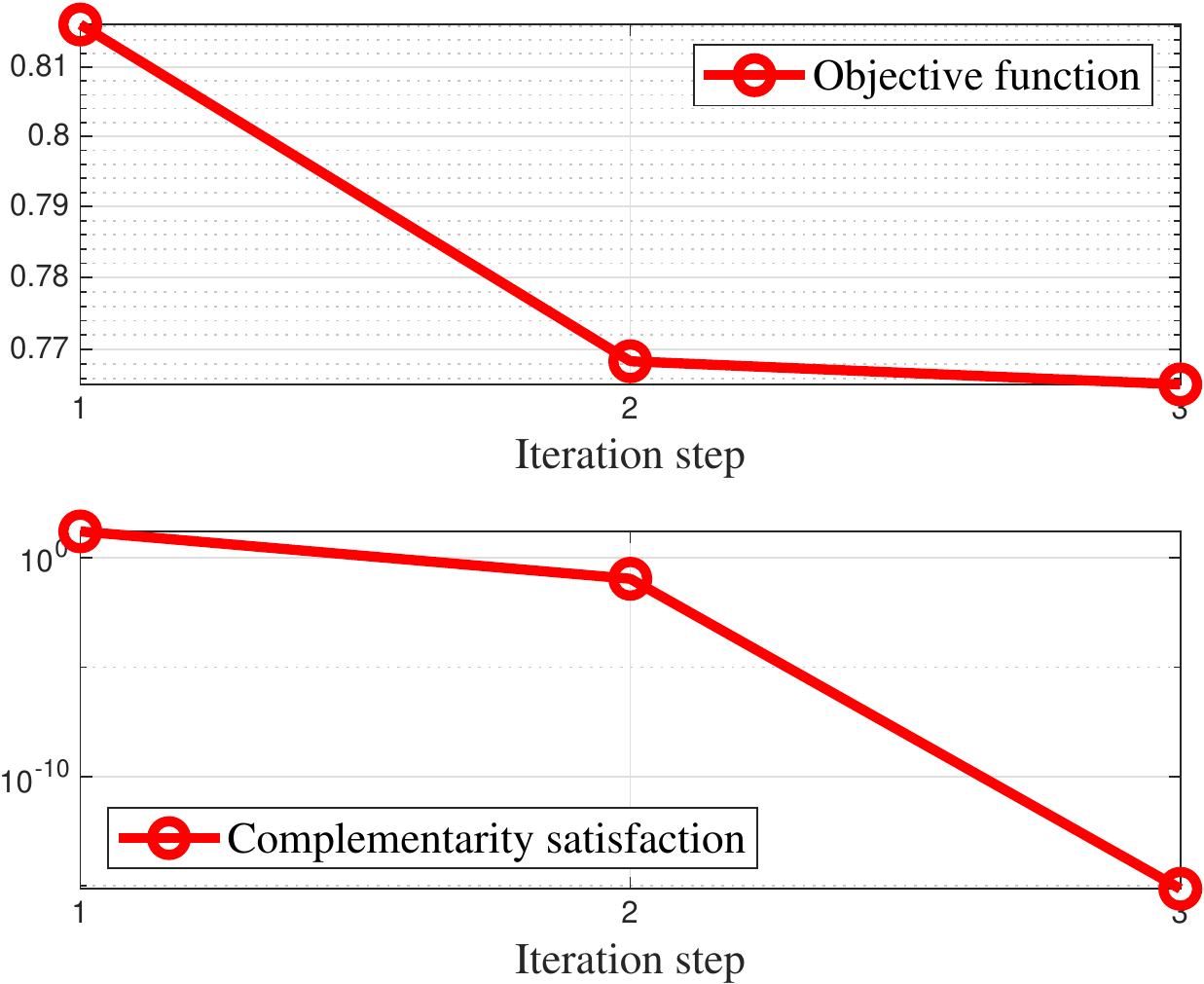}
	\caption{Objective function $\tilde{F}(x^*)$ and complementarity $|\varphi(x^*)|$ obtained by AD2 of \autoref{alg:AD-SLCQP}.}
	\label{fig:objfcnandcompslcqp}
\end{figure}

\autoref{fig:objfcnandcompslcqp} shows the convergence of objective function and complementarity satisfaction at AD2 step in the first iteration of \autoref{alg:AD-SLCQP}, and complementarity is well satisfied in two AD2 iterations.
\section{CONCLUSION}
This work proposed an algorithm to solve Mixed Boolean Nonlinear Programming in an antenna selection scenario. Our simulation of a $64\times 64$ antenna MU-MIMO network shows that, without destroying the BQP structure, AD-SBQP internally uses the linear approximation term of penalty function constructed from Boolean constraints, which results in extremely high complementary satisfaction accuracy compared with other algorithms. Moreover, the power consumption can be reduced to less than 55\% in several alternate direction steps.

%

\addtolength{\textheight}{-12cm}




\bibliographystyle{unsrt}
\bibliography{zsjcitation}
\end{document}